# Clustering into three groups on a quantum processor of five spins $S = 1$, controlled by pulses of resonant RF fields.

I. S. Pichkovskiy*. V. E. Zobov*.
*Kirensky Institute of Physics, Federal Research Center KSC SB RAS, Krasnoyarsk, Russia
rsa@iph.krasn.ru

**Abstract.**

We consider a quantum processor based on five qutrits represented by spins $S=1$, which is driven by radio frequency (RF) pulses selective in transitions between adjacent levels. Numerical simulation of the implementation of the quantum-adiabatic clustering algorithm was performed on the example of partitioning a set of six points into three groups. We find the amplitudes and durations of rectangular RF pulses, as well as the durations of free evolution intervals in the control pulse sequence, which made it possible to engineer a time-dependent effective Hamiltonian in the discrete-time approximation. Also we studied the dependence of the implementation fidelity on the parameters. We took quadrupole nuclei as qutrits, but the results obtained will be useful for controlling quantum processors based on qutrits represented by other systems.

## 1. Introduction

Multiple-pulse methods of magnetic resonance have made it possible to significantly increase the possibilities of this method in studying the local properties of substances [1]. They are based on the manipulation of the interactions between magnetic moments (spins) using a sequence of coherent pulses of radio frequency (RF) or microwave (MW) fields. In this way, an effective (average) Hamiltonian with given (required) characteristics is created [2–10].

Such effective Hamiltonians are currently used not only in spectroscopy, but also in solving problems in the field of quantum computing: modeling some quantum systems on others [11–13], studying the physics of quantum information [7, 14, 15], adiabatic quantum computing [16] (quantum annealing [17]), and many others. In this field, spin systems serve as a convenient object for studying regularities. Most of the work is done on spins S=1/2, which represent qubits (two-level quantum elements) [13, 18]. However, the interest of researchers in quantum computing on quantum elements with a large number of levels has recently grown [19]: with three levels, qutrits, or, in the general case, d levels, qudits. Such calculations have a number of advantages. They have greater noise immunity [20, 21]. Due to the use of additional levels, it is possible to more efficiently implement gates and algorithms on qubits [22–24]. Finally, when using $n$ qudits instead of qubits, the Hilbert space (computational basis) will grow by $(d/2)^n$ times [19, 25 - 27].

To date, several experimental implementations of coherent control of systems of qutrits have been performed: ions in a trap [28], quadrupole nitrogen nuclei in a crystal [5], NV centers in diamond [29–31], transmons in a superconductor [32, 33]. To control qutrits, pulses of an electromagnetic field (RF, MW or laser) are applied to the system, resonant to one or another pair of qutrit levels. These pulses cause selective state transformations of the selected two-level system, which can be represented by selective rotation operators. Sequences of selective rotation operators and the construction of an effective Hamiltonian for solving some problems on qudits were considered in several works: decoupling of the dipole-dipole interaction (DDI) [4, 5, 9, 10], factorization on two qudits [34] and three qutrits [35], modeling the propagation of quantum information in a system of five transmons [32].

Recently, we have shown theoretically that it is possible to solve artificial intelligence problems such as associative memory [36] or data clustering [37] using qutrits, by means of a slow (adiabatic) change of the Hamiltonian in time [16]. Then, in [38], we found a sequence of selective rotation operators to implement the data clustering algorithm. In the present paper, we will find a sequence of RF pulses to simulate the experimental solution of this problem on a quantum processor of five qutrits, represented by quadrupole nuclei with spins S=1, and coupled with DDI. After applying the found sequence to the system of spins, a set of six data points is grouped into three clusters according to the proximity of properties.

The article is structured as follows. In next section we consider the control Hamiltonians for qutrits. Section 3 describes a quantum processor based on 5 qutrits. In Section 4, we obtain a sequence of selective rotation operators for clustering using a simple example. In Section 5 we found a sequence of RF pulses. In Section 6, numerical simulations are performed. A brief summary is concluded in Section 7.

## 2. Adiabatic clustering algorithm on qutrits

We will study the control of the system of spins $S$=1, which represents a quantum processor on qutrits, using the example of clustering. The clustering task is to group data by proximity in the space of some properties [17]. We will consider points in a two-dimensional plane with coordinates ($x, y$) representing two properties of this data. The proximity of points with numbers $i$ and $j$ will be characterized by the Euclidean distance $R_{ij}$ between them:

$$R_{ij} = \sqrt{\left(x_i - x_j\right)^2 + \left(y_i - y_j\right)^2},  \qquad (1)$$

where ($x_i$, $y_i$) and ($x_j$, $y_j$) are coordinates of points $i$ and $j$ on the Cartesian plane. The solution of the clustering problem is to find such a partition of the set of $n$ points into $K$ clusters $C_\alpha$, which minimizes the sum of the sums of distances between points in each of the clusters:

$$W = \frac{1}{2}\sum_{\alpha=1}^{K} \sum_{i,j \in C_\alpha} R_{ij}. \qquad (2)$$

Let us consider case of partitioning a set of $n$ data points into three clusters, the simplest for qutrits. As a computational basis, we will use the basis $|m_1, m_2, ..., m_n\rangle$ of the eigenfunctions of the operator $S_i^z$ of spin projections onto the Z axis. For each data point $i$, we assign a qutrit represented by the spin operator $S_i^z$. The spin projection value

$m_i$, which takes one of three values: 1, 0, -1, denotes the belonging of point $i$ to one of the three clusters. The points with the same spin projections refer to the same cluster. The minimum value of the weight function (2) corresponds to the minimum value of the energy of the system with the Hamiltonian, which we proposed in [37],

$$H_f = \frac{1}{2}\sum_{i,j} H_{f\,ij} \tag{3}$$

$$H_{fij} = R_{ij}\left[S_i^z S_j^z + 3\left(S_i^z\right)^2\left(S_j^z\right)^2 - 2\left(S_i^z\right)^2 - 2\left(S_j^z\right)^2 + 1\right]. \tag{4}$$

Let us explain the choice of the form of the Hamiltonian (4). If the spins $i$ and $j$ refer to the same cluster and have the same projection values, then the expression in square brackets in (4) is equal to 1, and the contribution to the energy (3) takes a positive value $R_{ij}$. If the spins $i$ and $j$ refer to different clusters and have different projections, then the expression in square brackets in (4) is equal to -1, and the contribution to the energy (3) takes a negative value $-R_{ij}$. To obtain the minimum value of the total energy (3), the sum of positive contributions should be minimal (near spins), and the sum of negative contributions should be maximum (far spins).

We will solve the clustering problem via the slow (adiabatic) evolution of the system

$$\langle\Psi| = \langle\psi|\hat{Q}\exp\left(-i\int_0^T H(t)dt\right), \tag{5}$$

with the effective time-dependent Hamiltonian $0 \leq t \leq T$ [37]:

$$H(t) = \left(1 - \frac{t}{T}\right)H_0 + \frac{t}{T}H_f. \tag{6}$$

In (5) $\hat{Q}$ is time ordering operator, and in (6)

$$H_0 = -h\sum_{i=1}^n S_i^x \tag{7}$$

is the initial Hamiltonian of interaction with the transverse magnetic field ($S_i^x$ is the spin projection operator on the $X$ axis), the ground state $\langle\psi| = \frac{1}{2^n}\prod_{j=1}^n\left(|1\rangle_j + \sqrt{2}|0\rangle_j + |-1\rangle_j\right)$ of which can easily be prepared, and $H_f$ is the target Hamiltonian (3), the ground state $\langle\Psi|$ of which encodes the solution of our problem. At present, the quantum adiabatic clustering algorithm has already been implemented on qubits [17].

At the initial moment of time, the system of spins is prepared in the eigenstate of the Hamiltonian $H_0$ (7), which is a superposition of all variants of partitioning the set of points into three clusters. As a result of the adiabatic evolution (5) with Hamiltonian (6), the system will pass to the ground state of Hamiltonian (3), which corresponds to the minimum of the weight function (2) [17, 37]. The energy minimum determines the projections of the spin $m_i$ at the points. The points with same spin projections belong to the same cluster. Since in (4) the three projections 1, 0, -1 are equivalent, the ground state is sixfold degenerate. To remove the threefold degeneracy, we fix the value of the projection at one of the spins, for example, at the first spin $S_1^z = 1$. Thus, we have simplified the calculation by reducing the Hilbert space dimension by a factor of three and taking the Hamiltonian (3) in the following form:

$$H_f = \frac{1}{2}\sum_{i,j\neq 1} H_{f\,ij} + \sum_{j\neq 1} R_{1j}\left[S_j^z + \left(S_j^z\right)^2 - 1\right]. \tag{8}$$

## 3. Quantum processor on five spins.

In this section, we describe a five-spin quantum processor on which we will implement the proposed quantum algorithm. To provide addressing in control (selectivity), we take a system of five spins with different Larmor frequencies $\omega_i$ and different quadrupole constants $Q_i$ in a static external magnetic field and local crystal fields with a Hamiltonian (in frequency units):

$$H_5 = -\sum_{j=1}^5 \omega_j S_j^z + \sum_{i=j}^5 Q_j\left[3\left(S_j^z\right)^2 - 2\right] + H_{dd}. \tag{9}$$

The values $\omega_i$ and $Q_i$ are given in Table 1. Let us introduce the notation $E_j^1$, $E_j^2$ and $E_j^3$ for the energy levels of the spin $j$ with the projections $S_j^z$ 1, 0, -1, respectively. For the energy differences (transition frequencies) between adjacent levels of the spin $j$, we find

$$\omega_j^{1\leftrightarrow 2} = E_j^1 - E_j^2 = -3Q_j + \omega_j, \qquad \omega_j^{2\leftrightarrow 3} = E_j^2 - E_j^3 = 3Q_j + \omega_j. \tag{10}$$

$H_{dd}$ is Hamiltonian of the dipole-dipole interaction (DDI):

$$\begin{aligned}H_{dd} =\ & J_{12}S_1^z S_2^z + J_{13}S_1^z S_3^z + J_{14}S_1^z S_4^z + J_{15}S_1^z S_5^z + J_{23}S_2^z S_3^z + J_{24}S_2^z S_4^z + \\ & + J_{25}S_2^z S_5^z + J_{34}S_3^z S_4^z + J_{35}S_3^z S_5^z + J_{45}S_4^z S_5^z\end{aligned}, \tag{11}$$

in which we retained only the interaction between the longitudinal components of the spins (secular part), and neglected the interactions between the transverse components of the spins with different transition frequencies $\left|J_{ij}\right| \ll \left|\omega_i^{k\leftrightarrow n} - \omega_j^{p\leftrightarrow q}\right|$.

Table 1. Values of constants in Hamiltonian (9) used for calculations.

| Spin number | $\omega_i$ | $Q_i$ |
|---|---|---|
| 1 | 3000 | 15000 |
| 2 | 2500 | 10000 |
| 3 | 2800 | 12000 |
| 4 | 3200 | 18000 |
| 5 | 3800 | 30000 |

We will drive the state of the system using selective pulses of the RF magnetic field with frequencies equal to the selected transition between energy levels *k* and *n* of the system $\omega_{rf} = \omega_j^{k\leftrightarrow n}$ (10) [9, 34]. In a reference frame rotating with frequency $\omega_{rf}$ [39], the Hamiltonian of the action of a RF pulse takes the form:

$$H_{pulse} = -\sum_{j=1}^{5}\left(\omega_j - \omega_{rf}\right)S_j^z + \sum_{j=1}^{5}Q_j\left[3\left(S_j^z\right)^2 - 2\right] + H_{dd} + H_{field}, \tag{12}$$

where $H_{field}$ is Hamiltonian of interaction with a transverse RF magnetic field:

$$H_{field} = h_{pulse}\sum_{j=1}^{n}\left(S_j^x \cos\varphi - S_j^y \sin\varphi\right), \tag{13}$$

where $S_i^x$ and $S_i^y$ are spin projection operators on the *X* and *Y* axes, respectively, and $h_{pulse}$ is amplitude of the RF magnetic field (or RF pulse) in frequency units (in magnetic units the amplitude of the RF pulse is $h_{pulse}/\gamma_j$, where $\gamma_j$ - gyromagnetic ratio of spin *j*). In order to reduce crosstalk, this amplitude should satisfy the selectivity condition: Й. $\varphi$ is phase of the RF field of the applied pulse. For rotations around the *X* axis, we will take $\varphi = 0$ at a positive value of the angle of rotation and $\varphi = \pi$ at a negative value. For rotation around the *Y* axis, we will take $\varphi = \frac{3\pi}{2}$ and $\varphi = \frac{\pi}{2}$ respectively. To implement selective rotation by the angle $\Omega = \sqrt{2}h_{pulse}t_{pulse}$, the RF magnetic field $H_{field}$ (13) is switched on for a finite period of time $t_{pulse}$ ($t_{pulse} \gg 1/\omega_{rf}$). The action of the RF pulse on the state of the system is determined by the evolution operator:

$$\exp\left[-it_{pulse}H_{pulse}\right]. \tag{14}$$

In the ideal case, i. e. neglecting the DDI and the non-selective effect on other levels, the action of selective RF pulses (14) is represented by selective rotation operators $\{\Omega\}_{\alpha,j}^{k\leftrightarrow n}$ [9, 34, 35], which in matrix form look like:

$$\{\Omega\}_{z,j}^{1\leftrightarrow2}=\begin{pmatrix}\exp\left[-i\dfrac{\Omega}{2}\right]&0&0\\0&\exp\left[i\dfrac{\Omega}{2}\right]&0\\0&0&1\end{pmatrix},\{\Omega\}_{z,j}^{2\leftrightarrow3}=\begin{pmatrix}1&0&0\\0&\exp\left[-i\dfrac{\Omega}{2}\right]&0\\0&0&\exp\left[i\dfrac{\Omega}{2}\right]\end{pmatrix},$$

(15)

$$\{\Omega\}_{y,j}^{1\leftrightarrow2}=\begin{pmatrix}\cos\dfrac{\Omega}{2}&-\sin\dfrac{\Omega}{2}&0\\\sin\dfrac{\Omega}{2}&\cos\dfrac{\Omega}{2}&0\\0&0&1\end{pmatrix},\{\Omega\}_{y,j}^{2\leftrightarrow3}=\begin{pmatrix}1&0&0\\0&\cos\dfrac{\Omega}{2}&-\sin\dfrac{\Omega}{2}\\0&\sin\dfrac{\Omega}{2}&\cos\dfrac{\Omega}{2}\end{pmatrix},$$

where $\Omega$ is angle of rotation around the axis $\alpha$ ($\alpha = x, y, z$), $k$ and $n$ are the level numbers, $j$ is the number of rotating spin. The matrix of $X$ rotation differs from the matrix of $Y$ rotation by coefficients equal $(-i)$ in front of both sine functions.

Finally, for rotations around the Z axis, we will apply a composite RF pulse:

$$\{\theta\}_{z,i}^{k\leftrightarrow n}=\{-\pi/2\}_{y,i}^{k\leftrightarrow n}\cdot\{\theta\}_{x,i}^{k\leftrightarrow n}\cdot\{\pi/2\}_{y,i}^{k\leftrightarrow n}.$$

(16)

## 4. Creation of an effective interaction for partitioning a set of six points into three clusters.

Let us study the operation of a quantum processor on a simple example [37, 38]. Using a random number generator from the interval [-10, 10], we obtained the coordinates of six data points:

$$(4,-2), (-7,7), (6,-9), (-6,8), (-2,-6), (-9,5). \quad (17)$$

The first point (4,-2) has assigned the value of the projection $S^z = 1$ and the number 0. The rest of the points are assigned numbers from 1 to 5 in the order they follow. Having calculated the distances $R_{ij}$ (1) (given below in formulas (29) and (30)), we substitute them into the Hamiltonian (4) and (8). The result of clustering, performed in [37] using the proposed algorithm (5), is shown in Fig.1.

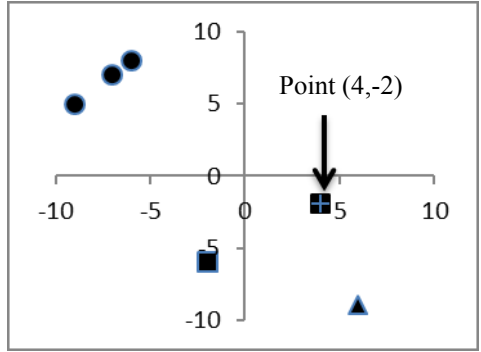

Fig.1. The result of partition of a set of six data points into three clusters. The clusters are marked by three types of markers: (-9.5), (-7.7) and (-6.8) by circle; (-2,-6) and (4,-2) by square; and (6,-9) by triangle.

Now let's consider the solution of this problem on the five-spin quantum processor (instead of the sixth spin, corresponding to the point (4,-2), terms representing the field from it are introduced into the second part of the Hamiltonian (8)). To implement algorithm (5), we should apply to the system a control sequence of selective RF pulses to engineer the time-dependent effective Hamiltonian (6) from the original Hamiltonian (9). In order to find the corresponding sequence of selective rotation operators, we pass in (5) to the discrete time $l$ ($0 \leq l \leq N$):

$$\langle\Psi|\approx\langle\psi|\prod_{l=0}^{N}U_l, \quad (18)$$

where

$$U_l=\exp\left\{-i\Delta t\left(\dfrac{l}{N}H_f+\left(1-\dfrac{l}{N}\right)H_0\right)\right\}\approx\exp\left[-i\dfrac{\Delta t}{2}\left(1-\dfrac{l}{N}\right)H_0\right]\exp\left[-i\Delta t H_f\dfrac{l}{N}\right]\exp\left[-i\dfrac{\Delta t}{2}\left(1-\dfrac{l}{N}\right)H_0\right]. \quad (19)$$

Here, following [16, 35 - 37], the operator of adiabatic evolution over time $T = \Delta tN$ with the Hamiltonian changing according to the linear law (6), we presented as a product of evolution operators on a sequence of $N$ small time intervals $\Delta t$. On each such interval, we will neglect the change in Hamiltonian (6). Then, we approximately represented the evolution operator (19) as a product of three non-commuting operators.

Let us substitute the Hamiltonians (8) and (4) into the evolution operator $\exp\left[-i\Delta t \dfrac{l}{N} H_f\right]$. We express it as a product of exponential operators corresponding to the separate terms of different types in $H_f$, which is possible, since all terms commute with each other.

First, we take the one-spin field terms. The corresponding evolution operators can be obtained by a sequence of operators of selective Z rotation as follows [34, 35]:

$$\exp\left[-i\Omega S_k^z\right] = \{2\Omega\}_{z,k}^{1\leftrightarrow 2} \{2\Omega\}_{z,k}^{2\leftrightarrow 3}. \tag{20}$$

The next one-spin terms in $H_f$ are quadratic in spin operators. The corresponding evolution operators can be obtained by the formula [34, 35]:

$$\exp\left[-i3\varphi\left(S_k^z\right)^2\right] = \{2\varphi\}_{z,k}^{1\leftrightarrow 2} \{-2\varphi\}_{z,k}^{2\leftrightarrow 3} \exp\left[-i2\varphi I\right], \tag{21}$$

where $I$ is identity matrix. For example:

$$\exp\left[-2i\Delta t_l R_{ij}(S_j^z)^2\right] = \{-4\Delta t_l R_{ij}/3\}_{z,j}^{1\leftrightarrow 2} \{4\Delta t_l R_{ij}/3\}_{z,j}^{2\leftrightarrow 3} \exp\left[4i\Delta t_l R_{ij} I/3\right] \tag{22}$$

Here and below, we have introduced the notation $\Delta t_l = \dfrac{l}{N}\Delta t$.

Now we consider the two-spin terms in $H_f$ and obtain the evolution operators $\exp\left[-i\Delta t_l 3R_{ij}\left(S_i^z\right)^2\left(S_j^z\right)^2\right]$ from the evolution operator under the action of DDI (11) $\exp\left[-it_d H_{dd}\right]$, where $t_d$ is the duration of the evolution interval, which will be determined later. First of all, we derive the factor of the desired form from the factor $\exp\left[-i\Delta t_l R_{ij} S_i^z S_j^z / 3\right]$ according to the following formula [34, 35]:

$$\begin{aligned}
\exp\left[-i\Delta t_l 3R_{ij}\left(S_i^z\right)^2\left(S_j^z\right)^2\right] &= \{4\Delta t_l R_{ij}/3\}_{z,i}^{1\leftrightarrow 2} \{-4\Delta t_l R_{ij}/3\}_{z,i}^{2\leftrightarrow 3} \times \\
&\times \exp\left[-4i\Delta t_l R_{ij} I/3\right] \times \{-\pi\}_{y,j}^{2\leftrightarrow 3} \{4\Delta t_l R_{ij}/3\}_{z,j}^{1\leftrightarrow 2} \{4\Delta t_l R_{ij}/3\}_{z,j}^{2\leftrightarrow 3} \times \\
&\times \{-\pi\}_{y,i}^{2\leftrightarrow 3} \exp\left[-i\Delta t_l R_{ij} S_i^z S_j^z / 3\right] \{-\pi\}_{y,i}^{1\leftrightarrow 2} \exp\left[-i\Delta t_l R_{ij} S_i^z S_j^z / 3\right] \times \\
&\times \{\pi\}_{y,i}^{1\leftrightarrow 2} \{\pi\}_{y,i}^{2\leftrightarrow 3} \{-\pi\}_{y,j}^{1\leftrightarrow 2} \{4\Delta t_l R_{ij}/3\}_{z,j}^{1\leftrightarrow 2} \{4\Delta t_l R_{ij}/3\}_{z,j}^{2\leftrightarrow 3} \{-\pi\}_{y,i}^{2\leftrightarrow 3} \times \\
&\times \exp\left[-i\Delta t_l R_{ij} S_i^z S_j^z / 3\right] \{-\pi\}_{y,i}^{1\leftrightarrow 2} \exp\left[-i\Delta t_l R_{ij} S_i^z S_j^z / 3\right] \{\pi\}_{y,i}^{1\leftrightarrow 2} \times \\
&\times \{\pi\}_{y,i}^{2\leftrightarrow 3} \{\pi\}_{y,j}^{1\leftrightarrow 2} \{\pi\}_{y,j}^{2\leftrightarrow 3}
\end{aligned} \tag{23}$$

This formula includes a factor containing only the DDI of a single pair of spins. To emphasize it from the general evolution of the system with Hamiltonian (11) containing the sum of all interactions, we divide the evolution interval into eight parts. Then we perform the inversion of individual spins on these intervals in such a way as to exclude unnecessary interactions. We obtain the spin inversion using the operators [34, 35]:

$$\begin{aligned}
P_k^{-1} &= \{-\pi\}_{y,k}^{1\leftrightarrow 2} \{-\pi\}_{y,k}^{2\leftrightarrow 3} \{-\pi\}_{y,k}^{1\leftrightarrow 2} \\
P_k &= \{\pi\}_{y,k}^{1\leftrightarrow 2} \{\pi\}_{y,k}^{2\leftrightarrow 3} \{\pi\}_{y,k}^{1\leftrightarrow 2} \\
P_k^{-1} S_k P_k &= -S_k
\end{aligned} \tag{24}$$

As an example, we get $\exp\left[-i8\Delta t_l J_{56} S_5^z S_6^z\right]$:

$$\begin{aligned}
\exp\left[-i8\Delta t_l J_{56} S_5^z S_6^z\right] &= P_1^{-1} \exp\left[-i\Delta t_l H_{dd}\right] P_1 \exp\left[-i\Delta t_l H_{dd}\right] \\
&\times P_2^{-1} P_1^{-1} \exp\left[-i\Delta t_l H_{dd}\right] P_1 \exp\left[-i\Delta t_l H_{dd}\right] P_2 \\
&\times P_3^{-1} P_1^{-1} \exp\left[-i\Delta t_l H_{dd}\right] P_1 \exp\left[-i\Delta t_l H_{dd}\right] \\
&\times P_2^{-1} P_1^{-1} \exp\left[-i\Delta t_l H_{dd}\right] P_1 \exp\left[-i\Delta t_l H_{dd}\right] P_2 P_3
\end{aligned} \tag{25}$$

In the evolution operators in (23), the corresponding exponents contains $\Delta t_l R_{ij}/3$, and in (25) it turned out $8t_d J_{ij}$. To bring it into conformity, we take the duration of the evolution interval with DDI as

$$t_d = \Delta t_l R_{ij} / (24 J_{ij}).  \qquad (26)$$

According to the rules described above, we have found the complete sequence of selective rotation operators and evolution intervals with the DDI Hamiltonian, which is necessary for clustering (18). In more detail, obtaining a sequence is considered in [38].

## 5. Parameters of RF pulses.

In this section, we implement selective rotation operators in the found sequence using rectangular RF pulses (14), which have three parameters: frequency $\omega_{rf}$, pulse duration $t_{pulse}$, and pulse amplitude $h_{pulse}$. The addressing of the selective rotation caused by the RF pulse is adjusted by selecting the RF pulse frequency equal to the desired transition frequency (10). To exclude the error caused by the phase shift from the energy levels of the main Hamiltonian, we take the duration $t_{pulse}$ as a multiple of all periods $\frac{2\pi}{\omega_j}$ and $\frac{2\pi}{Q_j}$ [40]. To this purpose, we have found that the constants given in Table 1 have the greatest common divisor $D = 100$. If we take

$$t_{pulse} = 2\pi \frac{C_j}{D},  \qquad (27)$$

where $C_j$ is an integer, then we will satisfy the required multiplicity condition for all constants. The choice of coefficients $C_j$ in specific cases will be discussed below. For given $t_{pulse}$ and angle of selective rotation $\Omega$, the amplitude of the RF pulse will be determined based on the following relation:

$$h_{pulse} = \frac{\Omega}{\sqrt{2} t_{pulse}} = \frac{D \Omega}{2\pi \sqrt{2} C_j}.  \qquad (28)$$

The DDI $H_{dd}$ (11) is also a source of error in (14). To reduce this error, we take the DDI constants much smaller than the amplitude of the pulse by introducing a small scale factor $\varepsilon = 10^{-6}$, while the ratios between the constants are determined by the distances (1) between points in our example (17) for clustering:

$$\begin{aligned}
H_{dd} = &-\frac{\sqrt{425}}{24}\varepsilon S_1^z S_2^z - \frac{\sqrt{2}}{24}\varepsilon S_1^z S_2^z - \frac{\sqrt{194}}{24}\varepsilon S_1^z S_1^z - \frac{\sqrt{8}}{24}\varepsilon S_1^z S_5^z - \\
&-\frac{\sqrt{433}}{24}\varepsilon S_2^z S_3^z - \frac{\sqrt{73}}{24}\varepsilon S_2^z S_4^z - \frac{\sqrt{421}}{24}\varepsilon S_2^z S_5^z - \frac{\sqrt{212}}{24}\varepsilon S_3^z S_4^z - \\
&-\frac{\sqrt{18}}{24}\varepsilon S_3^z S_5^z - \frac{\sqrt{170}}{24}\varepsilon S_4^z S_5^z
\end{aligned}  \qquad (29)$$

The small value of the DDI of quadrupole nuclei is related to their small dipole moments; this is a well-known fact [5, 41]. Additional adjustment of the values of the DDI constants (or rather their actions) can be performed using the durations of the time intervals, as we noted above (26). We take the interaction of five spins with the selected (excluded) spin included in the second term in Hamiltonian (8) in the following form:

$$\begin{aligned}
&\sqrt{202}\left[S_1^z + \left(S_1^z\right)^2 - 1\right] + \sqrt{53}\left[S_2^z + \left(S_2^z\right)^2 - 1\right] + \sqrt{200}\left[S_3^z + \left(S_3^z\right)^2 - 1\right] + \\
&+\sqrt{52}\left[S_4^z + \left(S_4^z\right)^2 - 1\right] + \sqrt{218}\left[S_5^z + \left(S_5^z\right)^2 - 1\right]
\end{aligned}  \qquad (30)$$

This part from (8) does not use the two-spin DDI in its construction and can be obtained using selective RF pulses (20) - (22).

Consider the specific RF pulses needed to solve the clustering problem. Let's start with the selective operators of rotation around the $Y$ axis by angles $\pi$, $-\pi$, $\frac{\pi}{2}$, $-\frac{\pi}{2}$, which are involved in the transformation of the factors of the evolution operator of the form (23) - (24) and (16). First, by sorting through the numbers $C_j$ in relation (28), we achieve the required value of the RF pulse amplitude $h_{pulse}$. Using this amplitude, on the basis of relation (27), we obtained the RF pulse duration $t_{pulse}$. The found numbers, amplitudes and pulse durations are shown in Table 2.

Table 2. Parameters of RF pulses for the implementation of rotations around the $Y$ axis by angles $\pi$ and $\frac{\pi}{2}$

| $\Omega$ | Spin number | $h_{pulse}$ | $t_{pulse}$ | $C_j$ |
|---|---|---|---|---|
| $\pi$ | 1 | 0.5124 | 4.3354 | 69 |
|  | 2 | 0.5124 | 4.3354 | 69 |
|  | 3 | 0.5124 | 4.3354 | 69 |
|  | 4 | 0.5277 | 4.2097 | 67 |
|  | 5 | 0.5277 | 4.2097 | 67 |
| $\dfrac{\pi}{2}$ | 1 | 0.2525 | 4.3982 | 70 |
|  | 2 | 0.2238 | 4.9637 | 79 |
|  | 3 | 0.26 | 4.2726 | 68 |
|  | 4 | 0.2525 | 4.3982 | 70 |
|  | 5 | 0.2996 | 3.7071 | 59 |

At the next stage, we consider the RF pulses of the rotation around the *X* axis by an arbitrary angle from relations (16) for *Z*-rotations in expressions (18), (22) and (23), which depend on the number *l*. First, at each step *l* we choose the integer $C_j$, for which we calculate the RF pulse amplitude based on relation (28). Then, using relation (27), we find the pulse duration. The parameters used in the program for these rotations are shown in Table 3. Since the angle, amplitude and duration of the pulse change at each step and are given by relations (27) and (28), we do not list them in the table 3.

Table 3. Parameters of RF pulses for the implementation of rotations around the *X* axis by angles $\Omega$, which change at each annealing step.

| $3\Omega$ | Spin number | $C_j$ |
|---|---|---|
| $4\Delta t_l \sqrt{170}$, $4\Delta t_l \sqrt{212}$, $4\Delta t_l \sqrt{73}$, $4\Delta t_l \sqrt{194}$, $6\Delta t_l \sqrt{52}$, $2\Delta t_l \left(\sqrt{52} - 2\sqrt{194} - 2\sqrt{73} - 2\sqrt{212} - 2\sqrt{170}\right)$ | 4 | 1 |
| $4\Delta t_l \sqrt{170}$, $4\Delta t_l \sqrt{18}$, $4\Delta t_l \sqrt{212}$, $4\Delta t_l \sqrt{8}$, $6\Delta t_l \sqrt{218}$, $2\Delta t_l \left(\sqrt{218} - 2\sqrt{8} - 2\sqrt{421} - 2\sqrt{18} - 2\sqrt{170}\right)$, | 5 | 1 |
| $4\Delta t_l \sqrt{18}$, $4\Delta t_l \sqrt{212}$, $4\Delta t_l \sqrt{433}$, $4\Delta t_l \sqrt{2}$, $6\Delta t_l \sqrt{200}$, $2\Delta t_l \left(\sqrt{200} - 2\sqrt{2} - 2\sqrt{433} - 2\sqrt{212} - 2\sqrt{18}\right)$, | 3 | 68 |
| $4\Delta t_l \sqrt{421}$, $4\Delta t_l \sqrt{73}$, $4\Delta t_l \sqrt{433}$, $4\Delta t_l \sqrt{425}$, $6\Delta t_l \sqrt{53}$, $2\Delta t_l \left(\sqrt{53} - 2\sqrt{425} - 2\sqrt{433} - 2\sqrt{73} - 2\sqrt{421}\right)$ | 2 | 80 |
| $4\Delta t_l \sqrt{8}$, $4\Delta t_l \sqrt{194}$, $4\Delta t_l \sqrt{2}$, $4\Delta t_l \sqrt{425}$, $2\Delta t_l \left(\sqrt{202} - 2\sqrt{425} - 2\sqrt{2} - 2\sqrt{194} - 2\sqrt{8}\right)$, $6\Delta t_l \sqrt{202}$ | 1 | 112 |

Now we obtain the factors $\exp\left[-i(\Delta t - \Delta t_l)H_0/2\right]$ in the evolution operator $U_l$ (19). Let us use the fact that the operators $S_j^x$ in $H_0$ commute with each other. This allows us to break the considered exponential operator

$$\exp\left[-i(\Delta t - \Delta t_l)H_0/2\right] = \prod_j \exp\left[-i(\Delta t - \Delta t_l)hS_j^x/2\right] \qquad (31)$$

to the product of five operators of rotations $\exp\left[-i(\Delta t - \Delta t_l)hS_j^x/2\right]$ of each spin *j* separately. We obtain each such operator by simultaneously acting on two transitions between levels $1 \leftrightarrow 2$ and $2 \leftrightarrow 3$ (10) by two RF fields [4, 9, 35] with frequencies $\omega_{1rf} = \omega_j^{1\leftrightarrow 2}$, $\omega_{2rf} = \omega_j^{2\leftrightarrow 3}$ and with equal amplitudes $h_{pulse}$ during the time $t_{pulse}$. To obtain the corresponding effective operator, we made a transformation to a generalized rotating frame [4, 42], which rotates not as usual with one frequency, but with two frequencies - at each transition with its own. After neglecting the rapidly oscillating contributions in the effective Hamiltonian, we obtain for it instead of (12) the following expression

$$H_{pulse} = -\sum_{j=1}^{5}\begin{pmatrix} \omega_j - \omega_{1rf} & 0 & 0 \\ 0 & 0 & 0 \\ 0 & 0 & -\omega_j + \omega_{2rf} \end{pmatrix}_j + \sum_{j=1}^{5} Q_j \left[ 3\left(S_j^z\right)^2 - 2 \right] + h_{pulse} \sum_{j=1}^{5} S_j^x + H_{dd}. \qquad (32)$$

We substitute this Hamiltonian into the RF pulse evolution operator (14) to perform the numerical calculation of the rotation of the spin $j$ specified by the operator $\exp\left[-i\Omega_2 S_j^x\right]$ in the product (31). The angle of rotation $\Omega_2 = t_{2pulse} h_{2pulse} = \left(1 - \dfrac{l}{N}\right)\dfrac{h\Delta t}{2}$, which is in the exponent, changes at each annealing step. We obtain the effective time $t_{2pulse}$ from relation (27), taking in this case an integer $C_j = 1$ for all five spins. From here we find the effective field. To calculate all five factors in (31), we use the same parameters $t_{2pulse}$ and $h_{2pulse}$. The frequencies (10) of the RF fields are different for different spins.

Finally, the factors $\exp\left[-it_d H_{dd}\right]$, included in (25), are obtained from the free evolution operator $\exp\left[-it_{free} H_5\right]$ with the total Hamiltonian (9) and with the duration of the time interval $t_{free}$, which differs from $t_d$ by a small value. We take time $t_{free}$ as a multiple of all periods $\dfrac{2\pi}{\omega_j}$ and $\dfrac{2\pi}{Q_j}$. As a result, we eliminate the contributions of $-\sum_{j=1}^{5}\omega_j S_j^z$ and $\sum_{j=1}^{5} Q_j \left(3\left[S_j^z\right]^2 - 2\right)$. For this purpose, we represent $\omega_j = D(\omega_j / D)$ and $Q_j = D(Q_j / D)$. Time $t_{free}$ is given by the equation

$$t_{free} \dfrac{\omega_j}{D} D = 2\pi C_0\left(\Delta t, N\right) l \dfrac{\omega_j}{D}, \qquad (33)$$

where $C_0\left(\Delta t, N\right)$ is an integer coefficient, which we will define as the nearest integer, after equating (33) to the exact phase shift $t_d \dfrac{\omega_j}{D} D$ over time interval $t_d$ (26). We obtain

$$C_0\left(\Delta t, N\right) = \text{closest integer to } \dfrac{\Delta t D}{2\pi N \varepsilon}. \qquad (34)$$

For example, for $N = 201$, $\Delta t = 0.052515$ and $\varepsilon = 10^{-6}$ we find $C_0 = 4158$, and $C_0 = 416$ for $\varepsilon = 10^{-5}$.

## 6. Computation and discussion

According to the rules derived in the previous section, a sequence of RF pulses and free evolution intervals was found to implement the clustering algorithm (18). The sequence contains $2369N$ pulses and $320N$ free evolution operators. A program for numerical simulation of the solution of the problem was written and calculations were performed for different values of the parameters $N$, $h$ and $\Delta t$. It should be noted that when passing from the representation (19) to the found sequence of RF pulses, these parameters are transformed into formal parameters in terms of which the rotation angles of selective operators are determined. For fixed values of the parameters $N$, $h$, and $\Delta t$ the physical parameters $h_{pulse}$, $t_{pulse}$ and $t_{free}$ change over a wide range from pulse to pulse in the sequence.

The result of calculation (18) is obtained in the form of a superposition of $3^5 = 243$ states of the computational basis (we fixed the selected spin in a state with a projection value of 1):

$$\langle \Psi | = \sum_{m_1, m_2, \ldots, m_5} C_{1, m_1, m_2, \ldots, m_5} \langle 1, m_1, m_2, \ldots, m_5 |. \qquad (35)$$

In the ideal case considered in [37], at the end of evolution at $t=T$, the system is in the state $\langle 1, -1, 0, -1, 1, -1 |$ with a probability of 0.99. The corresponding clustering result is shown in Fig. 1. The same result of clustering corresponds to the state $\langle 1, 0, -1, 0, 1, 0 |$ obtained after the rearrangement of the spin projections 0 and -1. At the end of evolution, the system is in this state with a probability of 0.01. Such a difference in probabilities was due to the fact that the curve for the instantaneous energy level [37] corresponding to this state passes above the curve corresponding to the state $\langle 1, -1, 0, -1, 1, -1 |$ throughout the evolution interval. The coincidence of the energies of the two states occurs only at $t=T$. Therefore, the probability of finding the system in this state is small. Symmetry breaking occurred due to fixing the value of the projection of the selected spin. On this basis, as the fidelity (accuracy) of the solution, we take the probability of finding the system in the state $|1, -1, 0, -1, 1, -1\rangle$:

$$F = |\langle \Psi | 1, -1, 0, -1, 1, -1 \rangle|^2 = |C_{1,-1,0,-1,1,-1}|^2. \qquad (36)$$

The dependences of the fidelity of the solution of the clustering problem on the parameters obtained as a result of numerical simulation are shown in Fig. 2-4. Figure 2 shows the dependence of fidelity on the duration of the discrete time step $\Delta t$. It can be seen that the fidelity monotonically increases up to the value of 0.05252, and then monotonically decreases. The decrease in fidelity at small values of $\Delta t$ is due to the violation of adiabaticity. The decrease in fidelity at large values of $\Delta t$ occurs due to the replacement of a continuous change in the Hamiltonian by a discrete one. The dependence of the fidelity on the magnitude of the transverse magnetic field $h$ shown in Figure 3 has a maximum at $h$=6.5. Figure 4 shows the dependence fidelity on the number of steps $N$. The fidelity increases monotonically with increasing number $N$, which indicates that the adiabatic condition is satisfied. At large values $N$, the increase of the fidelity stops at a certain limiting value, due to other contributions to the error. This error in Fig. 4 decreased with decreasing $\Delta t$ and increasing $h$. As a result, we brought the fidelity to the value of $F$=0.9887.

In addition, we have studied the dependence of fidelity on the value of the DDI. Some results of the simulation for different values of DDI are shown in Fig.4. A slight change in the fidelity with a tenfold increase in DDI indicates that the smallness of DDI, given in (29) by the scale factor $\varepsilon = 10^{-6}$, is taken with a large margin and that DDI of such a value makes an insignificant contribution to the error. We observed a noticeable decrease in fidelity to $F$=0.5 (compared to $F$=0.97) in our calculations with an increase in the scaling factor up to $\varepsilon = 9 \cdot 10^{-5}$ (at $N = 201$, $\Delta t = 0,05252$, $h = 6,5$). In the last example, when the field was increased to $h = 7.5$, the error decreased and the fidelity rose to $F = 0.79$, due to a decrease in the duration of the RF pulses. With a further increase in the amplitudes of the RF pulses, their selectivity will be violated, and the error will increase again. Such dependences of the error were studied in [34] on a simpler system of two qudits. On the other hand, in order to implement the algorithm with a small DDI, one has to take large durations of time intervals. This is acceptable in an isolated model system, but may lead to an increase in the error due to decoherence (relaxation) processes in real systems. The same applies to the long duration of selective RF pulses. To reduce their duration, experimenters use pulses of complex shape instead of rectangular ones [1, 2, 33, 40, 41, 43]. Finally, the number of RF pulses can be reduced if, instead of three pulses (16), in order to obtain a Z-rotation by an angle $\theta$, the phase of subsequent RF pulses acting on this transition is changed by the corresponding value [1, 32, 33].

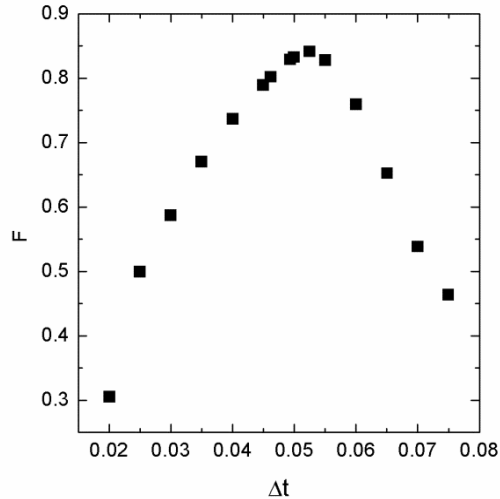

**Figure 2. The fidelity of clustering as a function of the discrete time step duration $\Delta t$ with $h = 4,3$, $N = 201$, $\varepsilon = 10^{-6}$ and with the parameters of selective RF pulses given in tables 2 - 3.**

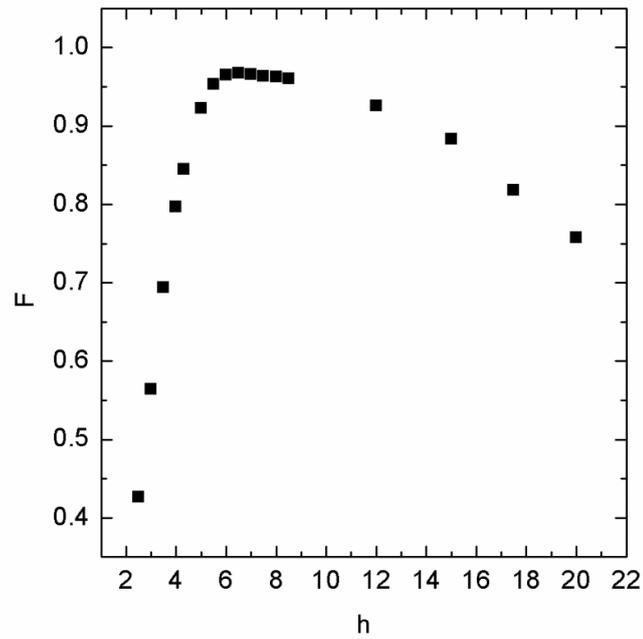

**Figure 3.** The fidelity of clustering as a function of the magnitude of the magnetic field $h$ with $N = 201$, $\Delta t = 0{,}05252$, $\varepsilon = 10^{-6}$ and with the parameters of selective RF pulses given in tables 2 – 3

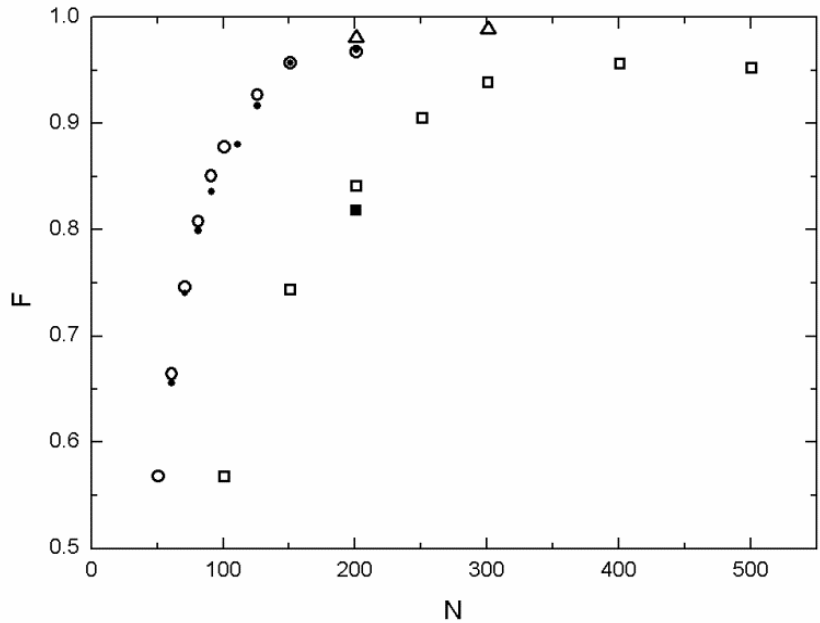

**Figure 4.** The fidelity of clustering as a function of the number of annealing steps $N$ for $\Delta t = 0{,}05252$ and with the parameters of selective RF pulses given in tables 2-3. The results obtained for the annealing field $h = 4{,}3$ are shown by empty squares, and for the field $h = 6{,}5$, by empty circles (triangles at $\Delta t = 0{,}0442$). Empty figures show the results of the calculation performed at $\varepsilon = 10^{-6}$. Filled figures show fidelity values obtained with the same parameters, but with $\varepsilon = 10^{-5}$.

# 7. Conclusion.

We have found a sequence of selective RF pulses and free evolution intervals to engineer a time-dependent control effective Hamiltonian for quantum annealing. As a result of the action of this sequence on a system of five spins $S$=1, we have solved the problem of partition of a set of 6 data points into three clusters. The dependence of the fidelity of obtaining the result on the physical parameters has been studied. The simulation has shown that the adiabatic clustering algorithm can be successfully performed on qutrits. Moreover, qutrits demonstrate advantages over qubits, since to solve the same problem on qubits, a system of 15 spins $S = 1/2$ is needed [17].

We simulated the operation of a five-qutrit quantum processor using five quadrupole nuclei with $S$=1 as an example. These can be nitrogen, lithium, or deuterium nuclei [5, 27, 41, 43]. However, the results obtained can be useful when using qutrits implemented on other quantum systems: trapped atoms and ions [19, 28, 44], superconducting systems [32, 33], objects with spin $S = 1$ in the magnetic and crystal fields. The latter include NV centers in diamond (paramagnetic color centers formed by electrons on vacancies near nitrogen atoms) [9, 10, 29–31].

## Acknowledgments

This study was supported by the Theoretical Physics and Mathematics Advancement Foundation "BA-SIS" #20-1-5-41-1. We are grateful for their trust and assistance in research.

## Declaration

**Competing interests:**

Author don't have competing interests

**Authors' contributions:**

Zobov V. E. wrote the main manuscript and Pichkovskiy I. S. prepared figures 1-4 and implement simulation. All authors reviewed the manuscript.

**Funding:**

This work funded by the Theoretical Physics and Mathematics Advancement Foundation "BA-SIS" #20-1-5-41-1.

**Availability of data and materials:**

Result of this work obtained via numerical simulation. All necessary parameters represented in section 5-6.

## References

[1] R. R. Ernst, G. Bodenhausen, A. Wokaun, Principles of nuclear magnetic resonance in one and two dimensions. (Oxford Univ. Press, Oxford, England, 1987)
[2] U. Haeberlen, High Resolution NMR in Solids. Selective Averaging (New York: Academic Press, 1976).
[3] B. N. Provotorov, E. B. Fel'dman, Sov. Phys. JETP 52, 1116 (1980)
[4] N. E. Ainbinder, G. B. Furman, Sov. Phys. JETP 58, 3, 575 (1983)
[5] D. Ya. Osokin, JETP 88, 5, 868 (1999)
[6] M. Bukov, L. D'Alessio A. Polkovnikov, Advances in Physics 64, 2, 139 (2015)
[7] E. I. Kuznetsova, E. B. Fel'dman D. E. Feldman, Physics-Uspekhi 59, 6, 577-582 (2016)
[8] A. Eckardt, Rev. Mod. Phys. 89, 1, 011004(2017).
[9] S. Choi, N. Y.Yao, M. D. Lukin, Phys. Rev. Lett. 119, 18, 183603 (2017)
[10] M. F. O'Keeffe, L. Horesh, J. F. Barry, D. A. Braje and I. L. Chuang, New Journal of Physics 21, 2, 023015 (2019).
[11] R. P. Feynman, Int. J. Theor. Physics 21, 467 (1982).
[12] I. M. Georgescu, S. Ashhab, F. Nori, Rev. Mod. Phys. 86, 1, 153 (2014)
[13] J. Preskill, Quantum 2, 79 (2018)
[14] G. A. Bochkin, S. G. Vasil'ev, S. I. Doronin, E. I. Kuznetsova, I. D. Lazarev, E. B. Fel'dman, Appl. Magn. Resonance 51, 7, 667 (2020)
[15] S. I. Doronin, E. B. Fel'dman, I. D. Lazarev, Phys. Let. A 406, 127458 (2021)
[16] T. Albash, D. A. Lidar, Rev. Mod. Phys. 90, 1, 015002 (2018).
[17] V. Kumar, G. Bass, C. Tomlin, J. Dulny, Quantum Inf. Process. 17, 39 (2018).
[18] M. A. Nielsen, I. L. Chuang, Quantum computation and quantum information (Cambridge University Press, Cambridge, England, 2000).
[19] Y. Wang, Z. Hu, B. C. Sanders, S. Kais, Frontiers in Physics 8, 479 (2020).
[20] D. Gottesman, A. Kitaev, J. Preskill, Phys. Rev. A 64, 1, 012310 (2001)
[21] E. T. Campbell, Phys. Rev. Lett. 113, 230501 (2014).
[22] T. C. Ralph, K. J. Resch, and A. Gilchrist, Phys. Rev. A 75, 022313 (2007).

[23] E. O. Kiktenko, A. S. Nikilaeva, Xu Peng, G. V. Shlyapnikov and A. K. Fedorov, Phys. Rev. A 101, 2, 022304 (2020)
[24] S. Bravyi, A. Kliesch, R. Koenig and E. Tang, Quantum 6, 678 (2022)
[25] A. Muthukrishnan, Jr C. R. Stroud, Phys. Rev. A 62, 052309 (2000)
[26] B. Tamir, Phys. Rev. A 77, 2, 022326 (2008)
[27] V. E. Zobov, D. I. Pekhterev, JETP letters 89, 5, 260(2009).
[28] C. Senko, P. Richerme, J. Smith, A. Lee, I. Cohen, A. Retzker and C. Monroe, Phys. Rev. X 5, 2, 021026 (2015).
[29] P. Neumann, R. Kolesov, B. Naydenov, J. Beck, F. Rempp, M. Steiner, V. Jacques, G. Balasubramanian, M.L. Markham, D.J. Twitchen, S. Pezzagna, J. Meijer, J. Twamley, F. Jelezko, J. Wrachtrup, Nature Physics 6, 249 - 253 (2010).
[30] J. Choi, S. Choi, G. Kucsko, P. C. Maurer, B. J. Shields, H. Sumiya, S. Onoda, J. Isoya, E. Demler, F. Jelezko, N.Y. Yao, and M. D. Lukin, Phys. Rev. Lett. 118, 093601 (2017).
[31] Z. Xu, Z. Q.Yin, Q. Han, T. Li, Optical Materials Express 9, 12, 465 (2019).
[32] M. S. Blok, V. V. Ramasesh, T. Schuster, K. O'Brien, J. M. Kreikebaum, D. Dahlen, A. Morvan, B. Yoshida, N. Y. Yao, and I. Siddiqi, Phys. Rev. X 11, 2. 021010 (2021)
[33] T. Roy, Z. Liy, E. Kapit, and D. I. Schuster, arXiv:2211.06523 [quant-ph] 12 Nov 2022
[34] V. E. Zobov, A. S. Ermilov, JETP 114, 6, 923(2012).
[35] V. E. Zobov, I. S. Pichkovskiy, Proc. SPIE (2018). https://doi.org/10.1117/12.2521253
[36] V. E. Zobov, I. S. Pichkovskiy, Quantum Inf. Process. 19, 342 (2020)
[37] V. E. Zobov, I. S. Pichkovskiy, Quantum Inf. Process. 21, 144 (2022)
[38] V. E. Zobov, I. S. Pichkovskiy, Proc. SPIE (2022). https://doi.org/10.1117/12.2622732
[39] C. P. Slichter, Principles of magnetic resonance, 2nd edition. (Springer-Veriag, 1980).
[40] R. Das, and A. Kumar: Phys. Rev. A **68**, 032304 (2003)
[41] R. Das, A. Mitra, S. V. Kumar, A. Kumar, International Journal of Quantum Information. 1, 3, 387-394 (2003)
[42] M. N. Leuenberger, D. Loss, Phys. Rev. B 68, 16, 165317 (2003)
[43] V. E. Zobov, V. P. Shauro, JETP 113, 2, 181 (2011)
[44] A. B. Klimov, R. Guzman, J. C. Retamal, C. Saavedra, Phys. Rev. A 67, 6, 062313 (2003)